\documentstyle[12pt]{article}
\begin{document}
\parskip=4pt
\parindent=18pt
\baselineskip=22pt
\setcounter{page}{1}
\centerline{\Large\bf An Exactly Solvable Model of Generalized Spin Ladder}
\vspace{6ex}
\parindent=18pt
\parskip=6pt
\begin{center}
{\large  Sergio Albeverio}$^\ast$\footnote{SFB 256; SFB 237; BiBoS; CERFIM
(Locarno); Acc.Arch., USI (Mendrisio)},
{\large  Shao-Ming Fei}$^\ast$\footnote{Institute of Physics,
Chinese Academy of Science, Beijing.} and
{\large Yupeng Wang}$^\dag$\footnote{AvH fellow.}
\end{center}
\begin{center}
\begin{minipage}{5.5in}
$^\ast$Institut f\"ur Angewandte Mathematik,
Universit\"at Bonn, D-53115 Bonn\par
\parindent=5pt
Fakult\"at f\"ur Mathematik, Ruhr-Universit\"at Bochum, D-4478 Bochum\\
$^\dag$Institut f\"ur Physik, Universit\"at Augsburg, 86135
Augsburg\par
Laboratory of Ultra-Low Temperature Physics, Chinese Academy of
Science,\par
Beijing 100080\\
\end{minipage}
\end{center}
\vskip 1 true cm

\begin{center}
\begin{minipage}{5in}
\vspace{3ex}
\centerline{\large Abstract}
\vspace{4ex}

A detailed study of an $S={1\over2}$ spin ladder model is given.
The ladder consists of plaquettes formed by
nearest neighbor rungs with all possible $SU(2)$-invariant interactions.
For properly chosen coupling constants, the model is shown to be
integrable in the sense that the quantum Yang-Baxter
equation holds and one has an infinite number of conserved quantities.
The $R$-matrix and $L$-operator associated with the
model Hamiltonian are given in a limiting case.  It is shown that after
a simple transformation, the model can be solved via a Bethe ansatz.
The phase diagram of the ground state is exactly derived using the Bethe 
ansatz equation.
\bigskip
\medskip

PACS numbers: 75.10.Jm

\end{minipage}
\end{center}

\newpage

Heisenberg spin ladders and generalized spin ladders
have attracted considerable attention in recent years, due to the
developing experimental results on ladder materials and
the hope to get some insight into the physics of metal-oxide
superconductors \cite{DagottoRice96}. Especially 
generalized ladders including other couplings
beyond the simplest case of rung and leg exchange
interpolate among a variety of systems and exhibit a
remarkably rich behavior [2-9]. In particular, it has been shown that the
diagonal interactions may cause frustration and change the
structure of the ground state \cite{BoseGayen93+,Weihong+},
while the biquadratic interactions, which 
can arise due to effective spin-spin interaction mediated
by phonons in real magnetic systems \cite{NersesyanTsvelik97},
tend to produce dimerization and may lead to a phase
transition into a ``non-Haldane'' spin liquid state with absence of
magnon excitations \cite{NersesyanTsvelik97,KM98prl}. 

As spin ladders are generally not equivalent to spin chains with nearest
neighbor interactions, till now little is known about integrable
spin ladder models\footnote{The present work was submitted for publication
in Euro. Phys. Lett.. During the preparation of a revised version
of this paper, another integrable ladder model without diagonal interactions
was presented in \cite{wang}.}. In this letter we
study a generalized $ S={1\over2}$ spin ladder system with both isotropic
exchange interactions and biquadratic interactions. Using ideas
related to the quantum Yang-Baxter
equations \cite{ybe} we found in our systems some cases of
integrable ladder systems, 
in the sense of models having an infinite number of conserved
quantities with explicit $R$ matrices satisfying the Yang-Baxter
equation. Properly choosing the spectral parameter, we get a Hamiltonian
consisting of only nearest-neighbor and next-nearest-neighbor interactions.
This model can be solved via an ordinary Bethe ansatz.

We consider a symmetric $16\times 16 $ matrix:
$$
\begin{array}{l}
\check{R}(x)=\\[4mm]
\left(
\begin{array}{cccccccccccccccc}
a_1 &&&&&&&&&&&&&&&\\[2mm]
&a_3&9a_2 &&3a_2 &&&&3b_2&&&&&&&\\[2mm]
&9a_2 &a_3&&3b_2&&&&3a_2 &&&&&&&\\[2mm]
&&&a_1 &&&&&&&&&&&&\\[2mm]
&3a_2 &3b_2&&a_5&&&&a_2&&&&&&&\\[2mm]
&&&&&16x&4a_2 &&&4b_2&2a_2 &&&&&\\[2mm]
&&&&&4a_2 &a_4&&&8a_2 &4b_2&&&&&\\[2mm]
&&&&&&&a_5&&&&a_2&&3a_2&3b_2&\\[2mm]
&3b_2&3a_2 &&a_2&&&&a_5&&&&&&&\\[2mm]
&&&&&4b_2&8a_2 &&&a_4&4a_2 &&&&&\\[2mm]
&&&&&2a_2 &4b_2&&&4a_2 &16x&&&&&\\[2mm]
&&&&&&&a_2&&&&a_5&&3b_2&3a_2 &\\[2mm]
&&&&&&&&&&&&a_1 &&&\\[2mm]
&&&&&&&3a_2 &&&&3b_2&&a_3&9a_2 &\\[2mm]
&&&&&&&3b_2&&&&3a_2 &&9a_2 &a_3&\\[2mm]
&&&&&&&&&&&&&&&a_1\end{array}
\right)
\end{array}
$$
where $a_1=2( -1 + 9x )$, $a_2=-b_2=( -1 + x )$, $a_3=7 + 9x$,
$a_4=2( 3 + 5x)$, $a_5=-1 + 17x$ and $x\in
{\ \hbox{\vrule width 0.6pt height 6.8pt depth 0pt \hskip -3.5 pt} C}$.
		      
Let $V$ denote a $4$-dimensional complex vector space.
$\check{R}(x)$ takes values in $End_{
{\ \hbox{\vrule width 0.6pt height 5.5pt depth 0pt
\hskip -2.8 pt} C}}(V\otimes V)$ and
satisfies the quantum Yang-Baxter equation \cite{ybe}:
\begin{equation}\label{ybe}
\check{R}_{12}(x)\check{R}_{23}(x y)\check{R}_{12}(y)
=\check{R}_{23}(y)\check{R}_{12}(x y)\check{R}_{23}(x),
\end{equation}
where $\check{R}_{ij}$ denotes the matrix on the complex vector space 
$V\otimes V\otimes V$, acting as $\check{R}(x)$ on the $i$-th and the 
$j$-th components and as the identity on the other components.
Namely $\check{R}_{12}=\check{R}\otimes{\bf 1}$,
$\check{R}_{23}={\bf 1}\otimes
\check{R}$ and ${\bf 1}$ is the identity operator on $V$. 

Let us set ${\cal H}=V_1\otimes V_2\otimes ...\otimes V_N$, $N\in {I\!\!
N}$. The corresponding $L$-operator acting on the $i$-th space $V_i$,
$i=1,2,...,N$, is a $4\times 4$ matrix with entries
$$
\begin{array}{rcl}
(L_i(x))_{11}&=&\frac{1}{2}(4 (3 x - 1)\sigma^0_{1,i}\sigma^0_{2,i}
+ c_1 (\sigma^0_{1,i}\sigma^z_{2,i} + \sigma^z_{1,i}\sigma^0_{2,i})\\[3mm]
&&+ 6 b_2 (\sigma^-_{1,i}\sigma^+_{2,i} + \sigma^+_{1,i}\sigma^-_{2,i})
+ 2 (1 + 3 x) \sigma^z_{1,i}\sigma^z_{2,i})\\[3mm]

(L_i(x))_{12}&=&\frac{1}{2}(
a_5 (\sigma^0_{1,i}\sigma^-_{2,i}
+ \sigma^z_{1,i}\sigma^-_{2,i})
+ a_2 (\sigma^-_{1,i}\sigma^0_{2,i} + \sigma^-_{1,i}\sigma^z_{2,i})
)\\[3mm]

(L_i(x))_{13}&=&\frac{1}{2}(
a_2 (\sigma^0_{1,i}\sigma^-_{2,i} + \sigma^z_{1,i}\sigma^-_{2,i})
+ a_5(\sigma^-_{1,i}\sigma^0_{2,i} + \sigma^-_{1,i}\sigma^z_{2,i})
       )\\[3mm]

(L_i(x))_{14}&=&2 c_1 \sigma^-_{1,i}\sigma^-_{2,i}\\[3mm] 

(L_i(x))_{21}&=&\frac{1}{2}(
a_3 (\sigma^0_{1,i}\sigma^+_{2,i} + \sigma^z_{1,i}\sigma^+_{2,i})
+ 9 a_2 (\sigma^+_{1,i}\sigma^0_{2,i} + \sigma^+_{1,i}\sigma^z_{2,i})
)\\[3mm]

(L_i(x))_{22}&=&\frac{1}{2}((15 x - 7)\sigma^0_{1,i}\sigma^0_{2,i}
       + 4 (1 + x)(\sigma^z_{1,i}\sigma^0_{2,i}
- \sigma^0_{1,i}\sigma^z_{2,i})\\[3mm]
&&+ 8 a_2 (\sigma^+_{1,i}\sigma^-_{2,i}
- \sigma^-_{1,i}\sigma^+_{2,i})
       - c_1 \sigma^z_{1,i}\sigma^z_{2,i})\\[3mm]

(L_i(x))_{23}&=&\frac{1}{2}(3 b_2(1 + \sigma^z_{1,i}\sigma^z_{2,i})
+ 2a_4 \sigma^-_{1,i}\sigma^+_{2,i}
+ 4 a_2 (\sigma^+_{1,i}\sigma^-_{2,i} + \sigma^0_{1,i}\sigma^z_{2,i}
       - \sigma^z_{1,i}\sigma^0_{2,i})
       )\\[3mm]

(L_i(x))_{24}&=&\frac{1}{2}(9 a_2 (\sigma^0_{1,i}\sigma^-_{2,i}
- \sigma^z_{1,i}\sigma^-_{2,i})
+ a_3 (\sigma^-_{1,i}\sigma^0_{2,i}-\sigma^-_{1,i}\sigma^z_{2,i})
              )\\[3mm] 

(L_i(x))_{31}&=&\frac{1}{2}(9 a_2 (\sigma^0_{1,i}\sigma^+_{2,i}
+\sigma^z_{1,i}\sigma^+_{2,i})
+ a_3 (\sigma^+_{1,i}\sigma^0_{2,i} + \sigma^+_{1,i}\sigma^z_{2,i})
              )\\[3mm]

(L_i(x))_{32}&=&\frac{1}{2}(3 b_2(1+ \sigma^z_{1,i}\sigma^z_{2,i})
+ 4 a_2 (\sigma^-_{1,i}\sigma^+_{2,i}
       + \sigma^z_{1,i}\sigma^0_{2,i} -
       \sigma^0_{1,i}\sigma^z_{2,i}) + 2a_4
       \sigma^+_{1,i}\sigma^-_{2,i}
       )\\[3mm]
       
(L_i(x))_{33}&=&\frac{1}{2}((15 x - 7)\sigma^0_{1,i}\sigma^0_{2,i}
+ 4 (1 + x) (\sigma^0_{1,i}\sigma^z_{2,i}
       - \sigma^z_{1,i}\sigma^0_{2,i} )\\[3mm]
&&+ 8 a_2 (\sigma^-_{1,i}\sigma^+_{2,i} - \sigma^+_{1,i}\sigma^-_{2,i})
       -c_1 \sigma^z_{1,i}\sigma^z_{2,i})\\[3mm]

(L_i(x))_{34}&=&\frac{1}{2}(a_3 (\sigma^0_{1,i}\sigma^-_{2,i}
- \sigma^z_{1,i}\sigma^-_{2,i})
       + 9 a_2 (\sigma^-_{1,i}\sigma^0_{2,i} -
        \sigma^-_{1,i}\sigma^z_{2,i})
       )\\[3mm] 

(L_i(x))_{41}&=&2 c_1 \sigma^+_{1,i}\sigma^+_{2,i}\\[3mm]

(L_i(x))_{42}&=&\frac{1}{2}(a_2 (\sigma^0_{1,i}\sigma^+_{2,i}
- \sigma^z_{1,i}\sigma^+_{2,i})
+ a_5 (\sigma^+_{1,i}\sigma^0_{2,i}
- \sigma^+_{1,i}\sigma^z_{2,i})
)\\[3mm]

(L_i(x))_{43}&=&\frac{1}{2}(a_5 (\sigma^0_{1,i}\sigma^+_{2,i}
- \sigma^z_{1,i}\sigma^+_{2,i})
+ a_2 (\sigma^+_{1,i}\sigma^0_{2,i}
- \sigma^+_{1,i}\sigma^z_{2,i})
)\\[3mm]
       
(L_i(x))_{44}&=&\frac{1}{2}(4(3 x - 1)\sigma^0_{1,i}\sigma^0_{2,i}
-c_1 (\sigma^0_{1,i}\sigma^z_{2,i}
+ \sigma^z_{1,i}\sigma^0_{2,i})\\[3mm]
&&+ 6 b_2 (\sigma^-_{1,i}\sigma^+_{2,i}
+ \sigma^+_{1,i}\sigma^-_{2,i})
+ 2 (1 + 3 x) \sigma^z_{1,i}\sigma^z_{2,i})
\end{array}
$$
where $c_1=a_1/2$, $\sigma^{\pm}_{\theta,i}$, $\sigma^{z}_{\theta,i}$
(resp. $\sigma^{0}_{\theta,i}$), $\theta=1,2$, are Pauli matrices
(resp. $2\times 2$ identity matrix) acting on the space $V_i$. A direct
calculation shows that $L_i(x)$ satisfies
\begin{equation}\label{rllllr}
\check{R}\left(\frac{x}{y}\right)[L_i(x)\otimes L_i(y)]
=[L_i(y)\otimes L_i(x)]\check{R}\left(\frac{x}{y}\right),~~~~~~i=1,2,...,N,
\end{equation}
where $x,y\in
{\ \hbox{\vrule width 0.6pt height 6.8pt depth 0pt
\hskip -3.5 pt} C}$, $y\neq 0$,
$\otimes$ is the tensor product of matrices. Let
\begin{equation}\label{T}
T(x)=L_N(x)L_{N-1}(x)...L_1(x).
\end{equation}
We have the fundamental commutation relations given by:
\begin{equation}\label{rttttr}
\check{R}\left(\frac{x}{y}\right)[T(x)\otimes T(y)]
=[T(y)\otimes T(x)]\check{R}\left(\frac{x}{y}\right).
\end{equation}

Let $t(x)=Tr_0 T(x)$, where $Tr_0$ takes trace on the space of $4\times
4$ matrix. According to (\ref{rttttr}) a system with Hamiltonian of the
form $H_{x_0}=J\frac{d}{d x}\log t(x)\vert_{x=x_0}$ for some $x_0\in
{\ \hbox{\vrule width 0.6pt height 6.8pt depth 0pt \hskip -3.5 pt} C}$
(such that $\log t(x)$ is defined and differentiable, e.g. $x_0=1$, see
(\ref{h})) and real constant $J$, acting on ${\cal H}$, has an infinite
number of conserved quantities $t(x)$:
\begin{equation}\label{Ht}
[H_{x_0},t(x)]=0, ~~~~~~\forall x\in
{\ \hbox{\vrule width 0.6pt height 6.8pt depth 0pt \hskip -3.5 pt} C},
\end{equation}
where $[\,,\,]$ stands for the commutator.
A system with Hamiltonian $H_{x_0}$ is then
by definition an integrable system. For arbitrary value of $x_0$,
$H_{x_0}$ generally describes integrable models
with long range interactions. These models can be exactly solved by
using the algebraic Bethe Ansatz method.

Let $P$ denote the permutation matrix on the ladder and set
$R(x)=P\check{R}(x)$. We see that $R(x)\vert_{x=1}$ is proportional to
the permutation matrix $P$. Therefore for $x_0=1$,
the Hamiltonian $H\equiv H_1$ describes a system with
nearest-neighbor interactions:
\begin{equation}\label{h}
\begin{array}{rcl}
H&=&\displaystyle\frac{d}{d x}\log t(x)\vert_{x=1}
\equiv \displaystyle\sum_{i=1}^{N}h_{i,i+1}\\[4mm]
&=&\displaystyle\sum_{i=1}^{N}
[5({\bf S}_{1,i}\cdot{\bf S}_{2,i} + {\bf S}_{1,i+1}\cdot{\bf
S}_{2,i+1})
+3({\bf S}_{1,i}\cdot{\bf S}_{1,i+1}+{\bf S}_{2,i}\cdot{\bf
S}_{2,i+1})\\[3mm]
&&- 3({\bf S}_{1,i}\cdot{\bf S}_{2,i+1}
+ {\bf S}_{2,i}\cdot{\bf S}_{1,i+1})
- 12({\bf S}_{2,i}\cdot{\bf S}_{1,i+1})({\bf S}_{1,i}\cdot{\bf
S}_{2,i+1})\\[3mm]
&&+ 20({\bf S}_{1,i}\cdot{\bf S}_{2,i})({\bf S}_{1,i+1}\cdot{\bf S}_{2,i+1})
+ 12({\bf S}_{1,i}\cdot{\bf S}_{1,i+1})({\bf S}_{2,i}\cdot{\bf S}_{2,i+1})
+\frac{57}{4}{\bf 1}\otimes{\bf 1}],
\end{array}
\end{equation}
where the periodic condition is assumed:
${\bf S}_{\theta,N+1}={\bf S}_{\theta,1}$, $\theta=1,2$ and
${\bf S}_{\theta,i}=(\sigma^x_{\theta,i},\sigma^y_{\theta,i},
\sigma^z_{\theta,i})/2$ is the spin vector operator on $V_i$.
Taking ${\bf S}_{1,i}$ (resp. ${\bf S}_{2,i}$) to be the spin operator on
the first (resp. second) leg of the $i$-th rung of a ladder,
and $V_i$ to be the tensor space for the action of these
two spin operators, the Hamiltonian (\ref{h})
describes an integrable spin ladder system with periodic boundary
conditions. The model is $SU(2)$-symmetric, i.e.,
$[H, {\cal S}^l]=0$, where
${\cal S}^l=\displaystyle\sum_{i=1}^N ({\bf S}^l_{1,i}+{\bf S}^l_{2,i})$,
$l=x,y,z$ are the total spin operators of the ladder.

Remark: Like the well known Affleck-Kennedy-Lieb-Tasaki model
\cite{AKLT}, the model (\ref{h}) derived from the transfer matrix
has no free parameters. Nevertheless,
as we shall show below, it can be generalized to a model
with two free parameters, i.e., the coupling constant along rungs and the
coupling constant of the rung-rung biquadratic interactions without losing
the integrability. The present model and the ones discussed in the
previous papers \cite{KM98prl,gsl} have same interaction terms in
Hamiltonian but with different coupling constants. The models in [5,9]
with more free parameters are generally not integrable.

To exactly solve the model (\ref{h}) the analytic algebraic Bethe Ansatz
method may be applied. The reference state with all the spins up is an
eigenstate. Some degenerate eigenstates
states can be obtained by applying the operator
${\cal S}^-={\cal S}^x-i{\cal S}^y$. The combinations
of the products of $(T(x))_{12}$
and $(T(x))_{13}$ can be used to construct ``Bethe Ansatz states" with an
arbitrary number of spins down. The relations of (\ref{rttttr}) would
then give the Bethe Ansatz equations. Some exact ground states can also be
constructed using the theorem in \cite{af}.

In the following we use a simpler method to solve the model.
Our Hamiltonian (\ref{h}) can be rewritten as (up to an irrelevant
constant term and a constant factor):
\begin{equation}\label{h1}
\begin{array}{rcl}
H&=&\displaystyle\frac12
\sum_{i=1}^N(\frac12+2{\bf S}_{1,i}\cdot{\bf S}_{1,i+1})
(\frac12+2{\bf S}_{2,i}\cdot{\bf S}_{2,i+1})
-\displaystyle\frac12\sum_{i=1}^N(\frac12+2{\bf S}_{1,i}\cdot{\bf S}_{2,i+1})
(\frac12+2{\bf S}_{2,i}\cdot{\bf S}_{1,i+1})\\[4mm]
&&+\displaystyle\frac{5}{6}\sum_{j=1}^N
(\frac12+2{\bf S}_{1,i}\cdot{\bf S}_{2,i})
(\frac12+2{\bf S}_{1,i+1}\cdot{\bf S}_{2,i+1}).
\end{array}
\end{equation}
We define the rung states as
$$
\begin{array}{l}
|0_i>=\displaystyle
\frac1{\sqrt2}(|\uparrow_{1,i},\downarrow_{2,i}>-|\downarrow_{1,i}
\uparrow_{2,i}>),~~~~~|1_i>=|\uparrow_{1,i},\uparrow_{2,i}>,\\[4mm]
|2_i>=\displaystyle
\frac1{\sqrt2}(|\uparrow_{1,i},\downarrow_{2,i}>+|\downarrow_{1,i},
\uparrow_{2,i}>),~~~~~|3_i>=|\downarrow_{1,i},\downarrow_{2,i}>,
\end{array}
$$
and the Hubbard operators
$X_i^{\alpha\beta}\equiv |\alpha_i><\beta_i|$, $\alpha$, $\beta=0,1,2,3$.
Hamiltonian (\ref{h1}) can be rewritten as
\begin{equation}\label{h2}
H=\frac12\sum_{i=1}^{N}\left[P_{i,i+1}^{rr}-P_{i,i+1}^{dd}
+\frac{20}3(\frac12-X_i^{00})(\frac12-X_{i+1}^{00})\right],
\end{equation}
where 
$$
P_{i,i+1}^{rr}=(\frac12+2{\bf S}_{1,i}\cdot{\bf S}_{1,i+1})
(\frac12+2{\bf S}_{2,i}\cdot{\bf S}_{2,i+1}),~~~~
P_{i,i+1}^{dd}=(\frac12+2{\bf S}_{1,i}\cdot{\bf S}_{2,i+1})
(\frac12+2{\bf S}_{2,i}\cdot{\bf S}_{1,i+1}).
$$

It can easily be checked that
$$
P_{i,i+1}^{rr}|\alpha_i>|\beta_{i+1}>=|\beta_i>|\alpha_{i+1}>,~~~~~
P_{i,i+1}^{dd}|\alpha_i>|\beta_{i+1}>=(-1)^{\epsilon(\alpha)+
\epsilon(\beta)}|\beta_i>|\alpha_{i+1}>,
$$
where $\epsilon(\alpha)$ is the parity of the state $|\alpha>$,
$\epsilon(0)=0$ and $\epsilon(1)=\epsilon(2)=\epsilon(3)=1$.
The $P$ operators can be expressed as
$$
P_{i,i+1}^{rr}=\sum_{\alpha,\beta=0}^3
X_i^{\alpha\beta}X_{i+1}^{\beta\alpha},~~~~~
P_{i,i+1}^{dd}=\sum_{\alpha,\beta=0}^3
(-1)^{\epsilon(\alpha)+\epsilon(\beta)}
X_i^{\alpha\beta}X_{i+1}^{\beta\alpha}.
$$
Therefore, we can rewrite the Hamiltonian (6) or (7) as
\begin{equation}\label{h12}
H=\sum_{i=1}^N\sum_{\alpha=1}^3 (X_i^{\alpha0}X_{i+1}^{0\alpha}
+X_i^{0\alpha}X_{i+1}^{\alpha0})+U\sum_{i=1}^N (\frac12-X_i^{00})
(\frac12-X_{i+1}^{00})-J\sum_{i=1}^N X_i^{00},
\end{equation}
with $U=\frac{10}3$, and $J=0$. In fact, the model (9) is
integrable for arbitrary real constants $U$ and $J$. This corresponds to
the case of an arbitrary rung exchange coupling
and an arbitrary rung-rung biquadratic coupling.
We shall discuss the generalized
integrable case rather than the special case (7) in the following text.

Obviously, $X_i^{\alpha\alpha}$ represents the number of the local state
$|\alpha_i>$ on the rung $i$ and satisfies the hard-core condition
$\sum_{\alpha=0}^3 X_i^{\alpha\alpha}=1$.
If we choose $|0>=|0_1>\otimes\cdots\otimes|0_N>$ as the vacuum state,
$X_i^{\alpha0}$ ($\alpha=1,2,3$) can be looked upon the creation
operators of $\alpha$-particles, i.e.,
$X_i^{\alpha0}|0_i>=|\alpha_i>$. The particle numbers
$N_\alpha=\sum_{i=1}^N X_i^{\alpha\alpha}$ are conserved quantities.
Notice that only three of them are independent since
$\sum_{\alpha=0}^3 N_\alpha=N$.
In this sense we construct the Bethe states
$|N_1,N_2,N_3>$. As shown in Eq.(9), there is no hybridization
among the states $|1>$, $|2>$, and $|3>$ because of the absence of
$X_i^{12}$, $X_i^{23}$, $X_i^{13}$ and their conjugates in Eq.(9). 
That means that the excitations from $|N_1, N_2,N_3>$ to $|N_1', N_2',N_3'>$
are dispersionless (i.e. have zero excitation energy)
as long as $N_1+N_2+N_3=N_1'+N_2'+N_3'$. In fact, the quantities
$$
Y_{21}=\sum_{i=1}^N X_i^{21}, {~~~~}Y_{31}=\sum_{i=1}^N X_i^{31},
$$
commute with the Hamiltonian, which means that the Bethe states are 
highly degenerate. The general eigenstates can be constructed
from $|N_e, 0,0>$:
$$
|N_1,N_2,N_3>=Y_{21}^{N_2}Y_{31}^{N_3}|N_e,0,0>,~~~~~
N_e=N_1+N_2+N_3.
$$
Therefore, we need only to consider the Bethe state $|N_e,0,0>\equiv|N_e>$.
This state reads:
$$
|N_e>=\sum_{n_1,\cdots,n_{N_e}}\Psi(n_1,n_2,\cdots,n_{N_e})\prod_{j=1}^{N_e}
X_{n_j}^{10}|0>,
$$
where $\Psi(n_1,n_2,\cdots,n_{N_e})$ is the wave function, and $n_j=1,\cdots,
N$ denotes the coordinate of the $j$-th triplet rung.
Let $\eta$ be defined by $U=2\cosh\eta$.
From an analysis similar to the one used in solving the $XXZ$ spin chain
(see, eg.\cite{14,15}), we have the Bethe Ansatz equation (for
$\lambda_j$, $j=1,..,N_e$)
\begin{equation}
\left[\frac{\sin(\lambda_j-\frac i2\eta)}
{\sin(\lambda_j+\frac i2\eta)}\right]^N
=-\prod_{l=1}^{N_e}\frac{\sin(\lambda_j-\lambda_l-i\eta)}
{\sin(\lambda_j-\lambda_l+i\eta)},
\end{equation}
and the eigenenergy to (\ref{h12})
(up to an irrelevant additive constant)
\begin{equation}
E=-\sum_{j=1}^{N_e}\left[\frac{2\sinh^2\eta}
{\cosh\eta-\cos2\lambda_j}-J\right],
\end{equation}
where $\lambda_j$ are the rapidities of the triplet rungs. We note that
a similar situation (namely mapping of a biquadratic spin-1 chain to an XXZ
Heisenberg chain) was discussed in [16].

The phase diagram of the ground state spanned by $J$ and $U$ is
almost the same to that of the $XXZ$ Heisenberg spin chain
with an effective magnetic field $J$, in the sense that
the triplet rungs and the singlet rungs
serve as the up spins and down spins, respectively.
We distinguish three regions,
according to $U>2$, $-2<U\leq 2$, $U\leq-2$ respectively:

\noindent (i) $U>2$:~
For $|J|<J_c$, the ground state is a Mott-like ``insulator" consisting of
$N/2$ triplet rungs and $N/2$ singlet rungs with an energy gap (a gap at
the lower end of the energy spectrum) $\Delta=J_c-|J|$, where $J_c$ is
given by
\begin{equation}
J_c=\frac{\pi\sinh\eta}\eta\sum_{n=-\infty}^\infty sech\frac{\pi^2}{2\eta}
(1+2n).
\end{equation}
For $J>U+2$, the triplet rungs are unfavorable and the ground state is a
rung-dimerized state (product of $N$ singlet rungs) with an energy gap 
$\Delta=J-(U+2)$, while for $J<-(U+2)$,
the ground state is a product of $N$ triplet rungs with an energy gap
$\Delta=|J|-(U+2)$. The latter two phases correspond to the
completely polarized states in the $XXZ$ spin chain. In the
intermediate parameter region $J_c\leq |J|
\leq U+2$ one has a gapless phase.

\noindent (ii) $-2<U\leq 2$:~ There is no Mott-like phase in this case.
For $J>U+2$, the ground state is still a rung-dimerized state
(consisting of only singlet rungs) and for $J<-(U+2)$, the ground state is
a product of $N$ triplet rungs. For $|J|\leq U+2$,
the ground state is a spin liquid with gapless spinon excitations
(cf. \cite{14} for a discussion of phenomena of this type).

\noindent (iii) $U\leq-2$:~ There is no gapless phase except
for $J=0$, $U=-2$. The ground state is almost the same as
that of a ferromagnetic spin chain (in the same sense as above).
For $J<0$, the ground state is a triplet-rung product while for
$J>0$ the ground state is a rung-dimerized state.

\vspace{2.5ex}
{\raggedleft Acknowledgements: We would like to thank Dr. R.H. Yue
(North-West University, Xian) for
very helpful discussions. The DFG and SFB-237 support to the second author
is gratefully acknowledged.}


\vspace{2ex}
\begin{thebibliography}{20}
\bibitem{DagottoRice96}  For a review see, e.g.,
E. Dagotto and T. M Rice, Science {\bf 271}, 618 (1996).

\bibitem{BoseGayen93+} I. Bose and S. Gayen, Phys. Rev. B {\bf 48},
10653 (1993);\\
Y. Xian, Phys. Rev. B {\bf 52}, 12485 (1995);

\bibitem{Weihong+} W.H. Zheng, V. Kotov, and J. Oitmaa, {\it Studies of
2-Chain Spin Ladder with Frustrating Second Neighbor Interactions},
cond-mat/9711006.

\bibitem{NersesyanTsvelik97} A. A Nersesyan and A. M. Tsvelik,
Phys. Rev. Lett. {\bf 78}, 3939 (1997).

\bibitem{KM98prl} A. K. Kolezhuk and H.-J. Mikeska, 
Phys. Rev. Lett. {\bf 80}, 2709 (1998).

\bibitem{BKMN98} S. Brehmer, A. K. Kolezhuk, H.-J. Mikeska and
U. Neugebauer, J. Phys.: Condens. Matter {\bf 10}, 1103 (1998).

\bibitem{KM97} A. K. Kolezhuk and H.-J. Mikeska, Phys. Rev. B {\bf
56}, R11 (1997).

\bibitem{BMN96} S. Brehmer, H.-J. Mikeska and U. Neugebauer,
J. Phys: Condens. Matter {\bf 8}, 7161 (1996).

\bibitem{gsl} A.K. Kolezhuk and H.-J. Mikeska, Int. J. Mod. Phys. B 
{\bf 12}, 2325 (1998).

\bibitem{ybe}
C.N. Yang, Phys. Rev. Lett. 19, 1312 (1967).\\
R.J. Baxter, {\it Exactly Solved Models in Statistical Physics},
Academic Press, New York 1982.

\bibitem{wang} Y. Wang, {\it Exact solution of a spin-ladder model},
cond-mat/9901168.

\bibitem{AKLT} I. Affleck, T. Kennedy, E H. Lieb and H. Tasaki,
Commun. Math. Phys. {\bf 115}, 477 (1988).

\bibitem{af}
S. Albeverio and S.M. Fei, {\sf EuroPhys. Lett.} {\bf 41}, 665 (1998).

\bibitem{14} C.N. Yang and C.P. Yang, Phys. Rev. {\bf 147},
303 (1966); {\bf 150},
321 (1966); {\bf 150}, 327 (1966); {\bf 151}, 258 (1966).

\bibitem{15} L.A. Takhatajian and L.D. Faddeev, Russ. Math. Surveys
{\bf 34:5}, 11 (1979) [Usp. Mat. Nauk. {\bf 34:5}, 13 (1979)].

\bibitem{16}
M.N. Barber and M.T. Batchelor, Phys. Rev. B {\bf 40}, 4621 (1989).

\end{thebibliography}
\end{document}